\documentclass[manuscript]{acmart}

\AtBeginDocument{%
  }

\setcopyright{acmlicensed}
\copyrightyear{2024}
\acmYear{2024}
\acmDOI{XXXXXXX.XXXXXXX}

\acmConference[FPT '24]{International Conference on Field Programmable Technology}{December 10--12,
  2024}{Sydney, Australia}
\acmISBN{978-1-4503-XXXX-X/18/06}

\acmJournal{TRETS} 
\acmVolume{37}
\acmNumber{4}
\acmArticle{111}
\acmMonth{8}

\usepackage{acronym}

\makeatletter
\newcommand*{\org@overidelabel}{}
\let\org@overridelabel\@verridelabel
\@ifpackagelater{acronym}{2015/03/21}{
	\renewcommand*{\@verridelabel}[1]{%
		\@bsphack
		\protected@write\@auxout{}{\string\AC@undonewlabel{#1@cref}}%
		\org@overridelabel{#1}%
		\@esphack
	}%
}{
	\renewcommand*{\@verridelabel}[1]{%
		\@bsphack
		\protected@write\@auxout{}{\string\undonewlabel{#1@cref}}%
		\org@overridelabel{#1}%
		\@esphack
	}%
}
\makeatother

\acrodef{ALAP}{as late as possible}
\acrodef{API}{application programming interface}
\acrodef{ASAP}{as soon as possible}
\acrodef{ASIC}{application-specific integrated circuit}
\acrodefindefinite{ASIC}{an}{an}
\acrodef{AXI}{advanced extensible interface}
\acrodef{BB}{basic block}
\acrodef{BE}{backend}
\acrodef{BLAS}{basic linear algebra subprograms}
\acrodef{BRAM}{block random access memory}
\acrodefplural{BRAM}{block random access memories}
\acrodef{BW}{bandwidth}
\acrodef{CC}{clock cycle}
\acrodef{CDC}{clock domain crossing}
\acrodef{CDFG}{control/dataflow graph}
\acrodef{CLB}{configurable logic block}
\acrodef{CNN}{convolutional neural network}
\acrodef{DDG}{data dependence graph}
\acrodef{DDR}{double-data-rate}
\acrodef{DNN}{deep neural network}
\acrodef{DSP}{digital signal processor}
\acrodef{DFG}{dataflow graph}
\acrodef{DSE}{design-space exploration}
\acrodef{DUT}{design under test}
\acrodefplural{DUT}{designs under test}
\acrodef{EDA}{electronic design automation}
\acrodef{FE}{frontend}
\acrodef{FF}{flip-flop}
\acrodef{FIFO}{first-in-first-out}
\acrodef{FPGA}{field-programmable gate array}
\acrodefindefinite{FPGA}{an}{a}
\acrodef{FPS}{frames per second}
\acrodef{FU}{functional unit}
\acrodef{GAT}{graph attention}
\acrodef{GEMV}{general matrix vector}
\acrodef{GNN}{graph neural network}
\acrodef{GSM}{global system for mobile communications}
\acrodef{HBC}{hybrid boundary condition}
\acrodef{HDL}{hardware description language}
\acrodef{HLS}{high-level synthesis}
\acrodefindefinite{HLS}{an}{an}
\acrodef{HW}{hardware}
\acrodef{II}{initiation interval}
\acrodefindefinite{II}{an}{an}
\acrodef{IP}{intellectual property}
\acrodefindefinite{IP}{an}{an}
\acrodef{IPI}{intellectual property integrator}
\acrodefindefinite{IPI}{an}{an}
\acrodef{IR}{intermediate representation}
\acrodefindefinite{IR}{an}{an}
\acrodef{LCS}{load-compute-store}
\acrodef{LSB}{least significant bit}
\acrodef{LUT}{look-up table}
\acrodef{MAC}{multiply-and-accumulate}
\acrodef{MAD}{multiply-and-add}
\acrodef{MIPS}{mega-instructions per second}
\acrodef{MCDFG}{multi-clock dataflow graph}
\acrodefindefinite{MCDFG}{an}{a}
\acrodef{MMM}{matrix-matrix multiplication}
\acrodef{MSB}{most significant bit}
\acrodef{MUL}{multiplier}
\acrodef{MVM}{matrix-vector multiplication}
\acrodef{OP}{operation}
\acrodef{PIPO}{ping-pong}
\acrodef{PPA}{power, performance, and area}
\acrodef{QoR}{quality of results}
\acrodef{RTL}{register-transfer level}
\acrodef{RTM}{reverse time migration}
\acrodef{SDC}{system of difference constraints}
\acrodef{SDFG}{synchronous dataflow graph}
\acrodef{SIMD}{single-instruction multiple-data}
\acrodef{SLP}{superword-level parallelism}
\acrodef{SoC}{system-on-chip}
\acrodef{SOTA}{state-of-the-art}
\acrodef{SRL}{shift-register lookup-table}
\acrodef{SCDFG}{single-clock dataflow graph}
\acrodefindefinite{SCDFG}{an}{a}
\acrodef{SNN}{spiking neural network}
\acrodef{SW}{software}
\acrodef{URAM}{ultra random access memory}
\acrodef{VMS}{virtual molecule screening}

\usepackage[inline]{enumitem}
\usepackage{multirow}
\usepackage{mathtools}
\usepackage[ruled, vlined]{algorithm2e}
\SetKw{Break}{break}
\SetKw{Continue}{continue}
\usepackage{subcaption}
\usepackage[capitalize]{cleveref}
\crefformat{equation}{(#2#1#3)}
\crefmultiformat{equation}{(#2#1#3)}{ and~(#2#1#3)}{, (#2#1#3)}{ and~(#2#1#3)}
\crefrangemultiformat{equation}{(#3#1#4)--(#5#2#6)}{,(#3#1#4)--(#5#2#6)}{,(#3#1#4)--(#5#2#6)}{,(#3#1#4)--(#5#2#6)}
\crefrangeformat{equation}{(#3#1#4)--(#5#2#6)}
\crefname{enumi}{}{}
\Crefname{enumi}{}{}

\usepackage{xcolor}

\definecolor{myred}{HTML}{D62728}
\definecolor{mydarkblue}{HTML}{1F77B4}
\definecolor{myviolet}{HTML}{9467BD}
\definecolor{myyellow}{RGB}{255,255,51}
\definecolor{mygrey}{HTML}{7F7F7F}
\definecolor{mybrown}{HTML}{8C564B}
\definecolor{mypink}{HTML}{E377C2}
\definecolor{lightyellow}{HTML}{faf1cb}
\definecolor{mygreen}{HTML}{6CA544}
\definecolor{mylightgreen}{HTML}{94B154}
\definecolor{mypurple}{HTML}{7F639E}
\definecolor{myblue}{HTML}{4F82BB}
\definecolor{mylightblue}{HTML}{4AA8C4}
\definecolor{myverylightblue}{HTML}{BCCFE0}
\definecolor{myorange}{HTML}{EA7A30}
\definecolor{mylightorange}{HTML}{F8B800}

\usepackage{siunitx}
\DeclareSIUnit{\one}{1}
\DeclareSIUnit{\frame}{f}

\iffalse
\colorlet{revadd}{mygreen}
\colorlet{revmod}{myorange}
\else
\colorlet{revadd}{black}
\colorlet{revmod}{black}
\fi

\begin{document}

\title[SILVIA: Automated Superword-Level Parallelism Exploitation via HLS-Specific LLVM Passes]{SILVIA: Automated Superword-Level Parallelism Exploitation via HLS-Specific LLVM Passes for Compute-Intensive FPGA Accelerators}

\author{Giovanni Brignone}
\email{giovanni.brignone@polito.it}
\affiliation{%
	\institution{Politecnico di Torino}
	\city{Turin}
	\country{Italy}
}
\author{Roberto Bosio}
\email{roberto_bosio@polito.it}
\affiliation{%
	\institution{Politecnico di Torino}
	\city{Turin}
	\country{Italy}
}
\author{Fabrizio Ottati}
\email{fabrizio.ottati@polito.it}
\affiliation{%
	\institution{Politecnico di Torino}
	\city{Turin}
	\country{Italy}
}
\author{Claudio Sansoè}
\email{claudio.sansoe@polito.it}
\affiliation{%
	\institution{Politecnico di Torino}
	\city{Turin}
	\country{Italy}
}
\author{Luciano Lavagno}
\email{luciano.lavagno@polito.it}
\affiliation{%
	\institution{Politecnico di Torino}
	\city{Turin}
	\country{Italy}
}


\begin{abstract}
	\Ac{HLS} aims at democratizing custom hardware acceleration with highly
	abstracted software-like descriptions.
	However, efficient accelerators still require substantial low-level hardware
	optimizations, defeating the \ac{HLS} intent.
	In the context of \aclp{FPGA}, \acp{DSP} are a crucial resource that typically
	requires a significant optimization effort for its efficient utilization,
	especially when used for sub-word vectorization.
	This work proposes SILVIA, an open-source LLVM transformation pass that
	automatically identifies superword-level parallelism within \iac{HLS}
	design and exploits it by packing multiple operations, such as additions,
	multiplications, and \aclp{MAD}, into a single \ac{DSP}.
	SILVIA is integrated in \textcolor{revadd}{the flow of} the commercial AMD Vitis HLS tool and proves its
	effectiveness by \textcolor{revmod}{packing multiple operations} on the \acp{DSP} without any manual
	source-code modifications on several diverse \acl{SOTA} \ac{HLS} designs such
	as \aclp{CNN} and \acl{BLAS} accelerators, reducing the \ac{DSP} utilization
	for additions by \SI{70}{\percent} and for multiplications and
	\aclp{MAD} by \SI{50}{\percent} on average.
\end{abstract}



\received{30 June 2024}
\received[revised]{30 September 2024}
\received[accepted]{13 November 2024}

\begin{teaserfigure}
	\subcaptionbox{\acs*{HLS} design source code.\label{subfig:c_source}}{
		\includegraphics{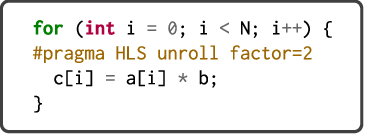}
	}
	\hfil
	\subcaptionbox{Original datapath.\label{subfig:sism_dsps}}[.25\linewidth]{
		\includegraphics{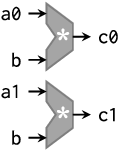}
	}
	\hfil
	\subcaptionbox{SILVIA datapath.\label{subfig:simd_dsp}}[.25\linewidth]{
		\includegraphics{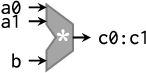}
	}
	\caption{Given a loop computing two multiplications in
	parallel~\subref{subfig:c_source}, the standard \acf*{HLS} flow
	generates the corresponding datapath \subref{subfig:sism_dsps} with two
	\acf*{DSP}, one per multiplication.
	The SILVIA flow automatically halves the \acs*{DSP} utilization by
	packing the two multiplications to a single \acs*{DSP}
	\subref{subfig:simd_dsp} without any manual source code modification.}
	\label{fig:example}
	\Description[Example of unoptimized and SILVIA-optimized datapaths]{
	A loop computing two multiplications with a shared operand allocates
	two DSPs with the standard HLS flow, while only one DSP with the
	proposed SILVIA HLS flow.
	}
\end{teaserfigure}
\maketitle

\acresetall

\section{Introduction}
Digital hardware design is a complex and time-consuming process, mostly
conducted at low abstraction levels using \acp{HDL}.
\Ac{HLS} aims to simplify this process by enabling the designers to describe
the hardware functionality via software programming languages, such as
\texttt{C} and \texttt{C++}.
The \ac{HLS} compiler translates the design description to a corresponding
\ac{RTL} implementation.
However, \ac{HLS} designs still need hardware-aware manual optimizations to
achieve high performance and efficiency.

In \ac{FPGA} platforms, low-precision data formats (e.g., 8-bit integers) are
crucial to achieve high computational intensity within the constraints of scarce
on-chip memory and limited off-chip memory bandwidth, particularly for \ac{SOTA}
machine learning applications, where 8-bit integers are the most popular
quantized data format for inference acceleration at the edge.
However, these data formats underutilize the large parallelism of the
\acp{FPGA}' \acp{DSP}, unless the designer explicitly packs the inputs of the
\acp{DSP} to efficiently utilize them.
For instance, the AMD/Xilinx UltraScale \ac{DSP}~\cite{dsp48} slice supports
the \ac{SIMD} operating mode for additions. Moreover, ingenious \ac{DSP}
packing approaches~\cite{squeezing,ssimd,acane,8bitpack,4bitpack,dacwinner}
enable other \ac{SIMD}\textcolor{revadd}{-like} operations such as multiple \acp{MAD} with shared
operands.

Several \ac{SOTA} \ac{FPGA} designs optimized for performance using \ac{DSP}
packing~\cite{minnella2023design,li2023firefly,zhang2022wsq,deepburning} prove the
effectiveness of this technique. However, current \ac{HLS} compilers and
\ac{HDL} synthesis tools lack automatic instruction vectorization capabilities
exploiting \ac{DSP} packing. Therefore, those designs required the manual
identification of the parallelism and the explicit fine-tuning of the source
code for taking advantage of the \textcolor{revmod}{operation-packing} capabilities of \acp{DSP}, relying on
either full \ac{RTL} implementations~\cite{li2023firefly}, or \ac{RTL}
implementation of the \ac{HLS} functions modeling the vectorized
operations~\cite{zhang2022wsq}, or mimicking the low-level \ac{RTL}
expressiveness in \ac{HLS} via bit operations to explicitly map the inputs and
outputs to the \ac{DSP} ports~\cite{minnella2023design,deepburning}, disrupting
the \ac{HLS} abstractions.

The automatic vectorization of loops and \acp{BB} is a well-established
optimization typically implemented in software compilers targeting
CPUs~\cite{autovectorizers,vectorization_gcc}.
However, since software compilers target inherently different hardware than
\ac{HLS}, they focus on issues not relevant for \ac{FPGA} designs, such as the
amortization of the overhead for moving data from scalar to vector register
files and vice versa.
Moreover, they only support basic \ac{SIMD} instructions operating on
independent data, missing more complex patterns such as two \acp{MAD} sharing
an operand~\cite{8bitpack}.

A multitude of studies~\cite{polyhedral,scalehls,soda_opt} automate code
transformations for improving \ac{QoR} of \ac{HLS} designs.
However, to the best of the authors' knowledge, no previous work automatically
identifies the compatible operations present in \ac{HLS} designs and packs them
to \acp{DSP} to improve the computational intensity.

SILVIA (automated \underline{S}uperword-level parallel\underline{I}sm
exploitation via \ac{HLS}-specific L\underline{LV}M passes for
compute-\underline{I}ntensive \ac{FPGA} \underline{A}ccelerators) extends the
\ac{SOTA} \ac{HLS} flow with additional compiler transformation passes
that automatically identify the compatible operations naturally
present in \ac{HLS} designs (e.g., exposed by loop unrolling) and map them into
\textcolor{revmod}{packed} \ac{DSP} operations, without any modification to the input
\texttt{C++} code.

For instance, the loop defined in \cref{subfig:c_source} computes two 8-bit
multiplications with a shared operand in parallel.
The \ac{SOTA} \ac{HLS} flow allocates two \acp{DSP}, one per multiplication
(\cref{subfig:sism_dsps}).
On the other hand, SILVIA automatically packs the two multiplications to a
single \ac{DSP}~\cite{8bitpack} (\cref{subfig:simd_dsp}) by analyzing and
transforming the \ac{IR} of the design.

\begin{figure}
	\centering
	\includegraphics{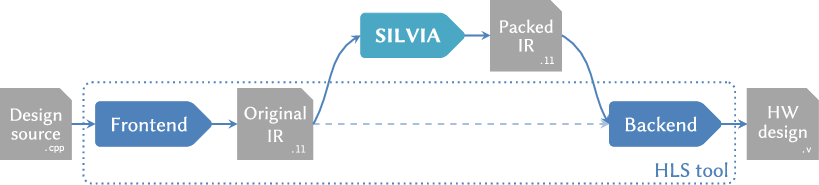}
	\caption{The modified \acl*{HLS} workflow with SILVIA. SILVIA optimizes
		the original LLVM \acl*{IR} generated by the \acl*{FE} for
		\textcolor{revmod}{\acs*{DSP}-packed} operations and provides
		it as input to the \acl*{BE}.}
	\label{fig:workflow}
	\Description[The SILVIA-optimized HLS flow]{
	The HLS frontend converts the design C/C++ source code to an LLVM IR.
	In the standard HLS flow this IR directly feeds the backend, which
	generates the corresponding non-packed hardware description.
	The SILVIA flow transforms the frontend-generated IR by packing its
	operations wherever possible.
	The backend converts this optimized IR to a DSP-packed hardware
	description.
	}
\end{figure}

The typical \ac{HLS} flow consists of the \ac{FE} step, that translates the
high-level design description into an optimized \ac{IR} (e.g., from
\texttt{C}/\texttt{C++} to LLVM \ac{IR}) and the \ac{BE} step, that generates
the hardware description of the functionality specified by the \ac{IR}.
SILVIA is executed between the \ac{FE} and the \ac{BE}, as shown in
\cref{fig:workflow}, because that point is accessible in most modern commercial and academic \ac{HLS} flows for \acp{FPGA}.

SILVIA is based on LLVM~\cite{llvm}, a widely used open-source compiler
infrastructure extensible via optimization passes that analyze and transform
the programs expressed in \iac{IR} form.
SILVIA is fully integrated in \textcolor{revadd}{the flow of} the commercial AMD Vitis HLS 2023.2 tool;
moreover, it could support any LLVM-based \ac{FPGA} \ac{HLS} tools (e.g.,
Dynamatic~\cite{dynamatic}, LegUp~\cite{canis2011legup}) with minor
adaptations.

The main contributions of this work are:
\begin{itemize}
    \item The open-source\footnotemark SILVIA framework, an LLVM
        transformation pass integrated in the AMD Vitis HLS \textcolor{revadd}{flow} that implements the
        generic functionality for packing multiple scalar operations within \ac{HLS}
	source code to single \acp{DSP} on \acp{FPGA} and is extensible to
	support different operations.
    \item The \texttt{SILVIAAdd} pass for binding two 24-bit or four 12-bit
	additions or subtractions to \iac{DSP} and the \texttt{SILVIAMuladd}
	for binding two 8-bit \acp{MAD} or four 4-bit multiplications with a
	shared operand to \iac{DSP}.
    \item The validation of SILVIA on several diverse designs, showing that it
        saves \SI{60}{\percent} addition \acp{DSP} and \SI{45}{\percent}
        multiplication and \ac{MAD} \ac{DSP} on average with no impact  on
        performance and no modification to the source code, compared to the
        original Vitis HLS synthesis flow.
        Moreover, it achieves results competitive with manually-optimized
        \ac{SOTA} \ac{CNN} accelerators.
\end{itemize}

\footnotetext{The SILVIA source code is available at
\url{github.com/brigio345/SILVIA}.}

{\color{revmod}
\section{Background}
}
SILVIA currently exploits the AMD/Xilinx UltraScale 48-bit~\cite{dsp48}
\textcolor{revadd}{and Versal 58-bit~\cite{dsp58}} \acp{DSP}
\textcolor{revmod}{operation-packing}
capabilities.
Among the multitude of proposed \ac{DSP} packing
methods~\cite{squeezing,ssimd,acane,8bitpack,4bitpack,dacwinner}, SILVIA
supports \textcolor{revmod}{packing} additions, subtractions,
multiplications, and \acp{MAD}.
Nevertheless, SILVIA is designed to be easily extended to support other \textcolor{revmod}{packed}
operations and \ac{DSP} architectures, as explained in \cref{sec:silvia}.


{\color{revmod}
\subsection{Additions and subtractions packing}\label{subsec:simd_add}
}
The \ac{DSP} architectures of the AMD/Xilinx UltraScale~\cite{dsp48}
\textcolor{revadd}{and Versal~\cite{dsp58}} \ac{FPGA} families support
\ac{SIMD} additions and subtractions.
Specifically, they can sum (or subtract) four independent pairs of signed or
unsigned operands on up to 12 bits (\texttt{four12} mode) or two independent
pairs of signed or unsigned operands on up to 24 bits (\texttt{two24} mode).
{\color{revmod}
\subsection{Factor-2 multiply-and-adds packing}\label{subsec:simd_madd}
}
Fu et al.\ \cite{8bitpack} proposed a methodology for computing two \acp{MAD}
of 8-bit operands, with one shared operand, on a single
UltraScale\textcolor{revadd}{/Versal} \ac{DSP}.

Specifically, if $a_i$ and $b_i$ are $m$-bits fixed-point numbers and $c_i$ is
an $n$-bits fixed-point number, $\forall i \in [1, N]$, $N$ \acp{DSP} can
compute
\begin{equation}
	\begin{array}{cc}
		p_a = \sum\limits_{i = 1}^{N} a_i \cdot c_i, &
		p_b = \sum\limits_{i = 1}^{N} b_i \cdot c_i.
	\end{array}
	\label{eqn:packing}
\end{equation}

The value of $p_a$ is mapped to the 30 \acp{MSB} and $p_b$ to the 18
\aclp*{LSB} of the \ac{DSP} output.
Therefore, to avoid $p_b$ overflowing into the $p_a$ bits,
\begin{equation}
	N \leq \begin{dcases}
		\left\lfloor \frac{2^{(18 - 1)}-1}{2^{(m - 1)}2^{(n - 1)}} \right\rfloor, & \textrm{if $b_i \cdot c_i$ is signed} \\
		\left\lfloor \frac{2^{18}-1}{(2^m-1)(2^n-1)} \right\rfloor,               & \textrm{otherwise}.
	\end{dcases}
	\label{eqn:mad_chain}
\end{equation}

For instance, with 8-bit signed operands, it is possible to chain of up to 7
\acp{DSP} computing \acp{MAD} without overflow.

It is worth noting that a single \ac{DSP} can compute two 8-bit multiplications
when $N = 1$.

{\color{revmod}
\subsection{Factor-4 multiplications packing}
}
The FINN framework~\cite{finn} provides an open-source implementation of the
architecture proposed by Preusser and Branca~\cite{preusser2020vectorization}
that multiplies four 4-bit signed factors by one common 4-bit factor (signed or
unsigned) using a single UltraScale\textcolor{revadd}{/Versal} \ac{DSP} and
some additional error-correction \ac{LUT} logic.
In particular, if $a_i$ are 4-bit signed fixed-point numbers and $b$ is a 4-bit
fixed-point number, their design computes
\begin{equation}
p_i = a_i \cdot b, \forall i \in [0, 3].
\label{eqn:4b_mul}
\end{equation}

In the context of \ac{CNN} accelerators~\cite{dnn_survey}, this packing is
particular effective for the feature map reuse (i.e., the same activation
multiplied by different signed weights).
However, it does not support the filter reuse (i.e., the same weight multiplied
by different activations, that are unsigned when the activation function is a
rectified linear unit).

\begin{figure}
\subcaptionbox{\acs*{DSP} operations.\label{subfig:4b_dsp}}{
	\includegraphics[scale=.55]{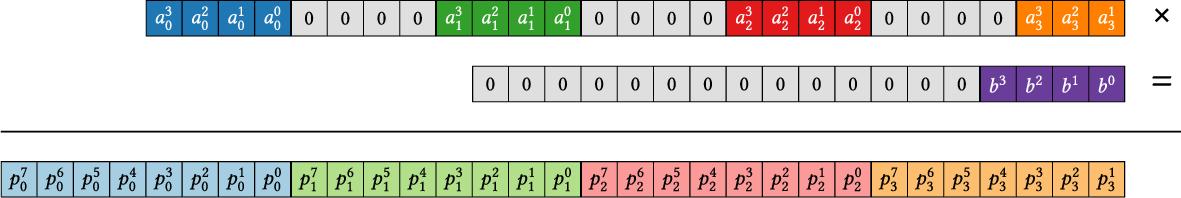}
}
\hfil
\subcaptionbox{\acsp*{LUT} operations.\label{subfig:4b_lut}}{
	\includegraphics[scale=.55]{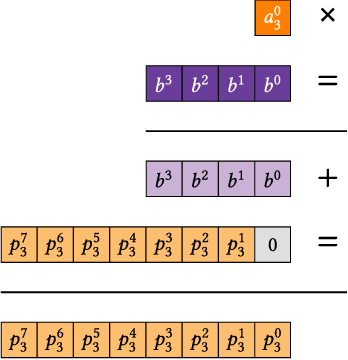}
}
\caption{The bit mapping of the proposed method for computing four
multiplications between four 4-bit unsigned factors and one common 4-bit factor
(signed or unsigned).}
\label{fig:4b_pack}
\Description[DSP packing for 4-bit mutiplication]{
The 27-bit DSP input accommodates three factors and the three MSBs of the
fourth factor, with four zero-padding bits in between each factor.
The 18-bit DSP input accommodates the common factor.
The DSP output contains the products of the first three factors by the common
factor, and the partial result of the fourth factor by the common factor.
To obtain the fourth product, the LSB of the fourth factor is multiplied by the
common factor and is added to the partial result.
}
\end{figure}

Therefore, this work supports \ac{DSP} packing for the filter reuse too, by
introducing a novel packing mechanism to multiply four 4-bit unsigned factors by
one common 4-bit factor (signed or unsigned) with a single
UltraScale\textcolor{revadd}{/Versal} \ac{DSP} and a small amount of \acp{LUT}
as shown by \cref{fig:4b_pack}.

Specifically, a multiplication between two 4-bit values requires 8 bits of
output precision to avoid overflow.
{\color{revmod}The UltraScale and the Versal \acp{DSP} provide 27\texttimes18
and 27\texttimes24 multipliers, respectively.
\Cref{subfig:4b_dsp} shows that the proposed packing maps three 4-bit operands
interleaved with four zeros of padding (to reserve eight output bits and avoid
overflow) and the three \acp{MSB} of the fourth operand $a_3$ to the 27-bit
input, while the other input (i.e., the 18 or 24-bit port) accommodates the
common operand $b$.}
A second step, implemented in \acp{LUT} and depicted in \cref{subfig:4b_lut},
computes the final product $p_3$ according to
\begin{equation}
p_3 = a_3 \cdot b = ((a_3^{[3:1]} \cdot b) \cdot 2) + (a_3^0 \cdot b).
\end{equation}
The \ac{DSP} calculates the most expensive portion of the computation (i.e.,
$a_3^{[3:1]} \cdot b$).
The multiplication by 2 does not require any additional hardware, since it is a
left-shift by one position.
The multiplication of $b$ by one bit $a_0^0$ is hardware-friendly too (e.g.,
it can be implemented by the \emph{and} operation between $a_0^0$ and each bit
of $b$).
Finally, the addition requires a small adder with 4-bit and 8-bit operands.

If the common operand $b$ is signed, the products $p_i$ must be corrected
similarly to the method by Fu et al.\ \cite{8bitpack} (i.e., adding the
\ac{MSB} of a product $p_i$ to the next product $p_{i + 1}$).

{\color{revmod}
\section{Methodology}\label{sec:silvia}
}
The SILVIA LLVM optimization pass is executed in the middle of the typical
LLVM-based \ac{HLS} flow (e.g., Vitis HLS), by optimizing for
\textcolor{revmod}{\ac{DSP}-packed} operations the \ac{FE}-generated
LLVM~\cite{llvm} \ac{IR} before passing it to the \ac{BE}, as shown in
\cref{fig:workflow}.
With this approach, SILVIA processes \iac{IR} that is already optimized by the
\ac{FE} (e.g., the dead code is eliminated; the width of the instructions is
minimized) and its high abstraction level allows for using the advanced
analysis and transformation facilities provided by the LLVM \acp{API}, before
it is lowered to hardware description by the \ac{BE}.

SILVIA is integrated in the Vitis HLS flow. Therefore, it complies with the
LLVM 3.1~\cite{llvm} \acp{API} for compatibility with the \ac{FE}-generated
\ac{IR}.
The SILVIA flow does not exploit the Vitis HLS capability of inserting
user-defined passes within the \ac{FE} itself~\cite{xilinx2024HLS}, because this
approach would execute SILVIA early in the \ac{FE} pipeline, preventing it from
taking advantage of the \ac{FE} optimizations, such as the width minimization of
the instructions.

\begin{algorithm}
	\DontPrintSemicolon
	\KwData{$BB$ = basic block belonging to \iac{HLS} design.}
	\KwResult{$BB^*$ = basic block functionally equivalent to $BB$ and
		optimized for \acs*{DSP}-packed operations.}

	$C \gets \textcolor{myblue}{\text{\texttt{getCandidates}}}(BB)$\;

	$BB^* \gets BB$\;
	\tcp{Maximize the space for valid tuples.}
	\For{$c \in C$}{
		$BB^* \gets \text{\texttt{moveUsesALAP}}(c, BB^*)$\;
	}

	\tcp{Group the candidates in valid tuples.}
	$\mathcal{T} \gets \textcolor{mygreen}{\text{\texttt{getTuples}}}(C)$\;

	\tcp{Pack the valid tuples.}
	\For{$T \in \mathcal{T}$}{
		$P \gets \textcolor{myblue}{\text{\texttt{packTuple}}}(T)$\;
		$BB^* \gets \text{\texttt{replaceTuple}}(T, P, BB^*)$\;
	}

	\caption{SILVIA's main optimization routine.}\label{alg:SILVIA}
\end{algorithm}

\Cref{alg:SILVIA}
summarizes the main steps of the SILVIA optimization pass.
SILVIA optimizes one \ac{BB} at a time, similarly to the
superword-level vectorizers targeting CPUs~\cite{autovectorizers}.
It collects the \emph{candidate} instructions amenable for vectorization,
groups them into \emph{tuples} of compatible candidates, and finally replaces
each tuple with an optimized \emph{packed operation}.
A concrete example is the automatic binding of four 12-bit additions to a
single \ac{DSP} configured in the \texttt{four12} mode.
SILVIA searches for 12-bit \texttt{add} instruction candidates, groups them
into tuples of four elements, and binds each tuple to a single \ac{DSP}.

The \texttt{SILVIA} class extends the LLVM \texttt{BasicBlockPass} class and
implements the structure of \cref{alg:SILVIA}.
The structure of the \texttt{SILVIA} class allows supporting different
\textcolor{revmod}{\ac{DSP}-packed} operations through derived classes of
\texttt{SILVIA}, which simply override some virtual functions of
\texttt{SILVIA} and exploit the rest of the existing framework that is common
to every \textcolor{revmod}{packed} operation.
Specifically, the derived classes must override the \texttt{getCandidates} and
\texttt{packTuple} functions (highlighted in blue in \cref{alg:SILVIA}), and
the \texttt{canPack} and \texttt{isTupleFull} functions, used internally by
\texttt{getTuples}.

The \texttt{SILVIA} class is currently extended by two examples of \ac{DSP}
packing specializations:
\begin{itemize}
	\item \texttt{SILVIAAdd}: four 12-bit additions (or subtractions) or
		two 24-bit additions (or subtractions), discussed in
		\cref{subsec:simd_add}.
	\item \texttt{SILVIAMuladd}: two 8-bit \acp{MAD} (or multiplications)
		or four 4-bit multiplications, discussed in
		\cref{subsec:simd_madd}.
\end{itemize}

\subsection{Candidate identification}
The \texttt{getCandidates} function identifies the candidates as the initial
step of the SILVIA flow.
Given \iac{BB}, \texttt{getCandidates} returns the set of instructions (or
patterns of instructions) amenable for the packing optimization.

For the \texttt{SILVIAAdd} pass, \texttt{getCandidates}
returns the addition instructions whose operands size in
bits is within the allowed range (up to 12 or 24 bits).


The \texttt{getCandidate} of \texttt{SILVIAMuladd}, instead, searches for trees
of addition instructions whose leaves are multiplication instructions between
operands of 8-bit or less, for the factor-2 packing, or 4-bit or less, for the
factor-4 packing.
It is worth noting that the \texttt{SILVIAMuladd} pass also supports
multiplication-only packing, since a degenerate tree composed of a single
multiplication is a valid candidate too.


\subsection{Tuple generation}
%
Given a set of candidates, the \texttt{getTuples} function combines them into
tuples
\begin{enumerate}[label=(\alph*)]
	\item\label{it:data_deps} whose candidates do not depend on each other,
	\item\label{it:insert_point} with an available insertion point between
		the first use of each candidate and after the last definition
		of each candidate's operands,
	\item\label{it:can_pack} that satisfy the constraints of a specific
	\textcolor{revmod}{\ac{DSP}-packed} operation.
\end{enumerate}

%

\subsubsection{Pack insertion point}\label{subsubsec:insert_point}
SILVIA replaces a tuple with a packed operation by inserting a call to a
function that implements the packed operation (further details in
\cref{subsec:packing}).
The packed function call must be placed after the definition of every tuple's
operand and before every use of the tuple's results to produce valid LLVM code.
However, when compiling \texttt{C} code containing unrolled loops (a typical
source of parallelism and potential vectorization), the LLVM compiler inserts
multiple copies of the loop body in sequence.

\begin{figure}
	\subcaptionbox{Source code.\label{subfig:llvm_source}}{
		\includegraphics{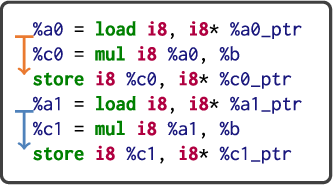}
	}
	\hfil
	\subcaptionbox{\acs*{ALAP}-rearranged code.\label{subfig:llvm_rearrange}}{
		\includegraphics{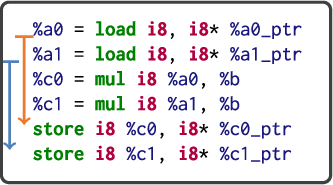}
	}

	\subcaptionbox{\acs*{DSP}-packed code.\label{subfig:llvm_packed}}{
		\includegraphics{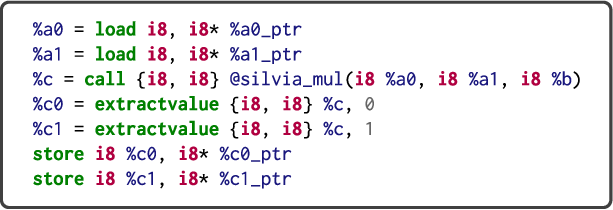}
	}
	\caption{The \texttt{C} code defined in \cref{subfig:c_source} compiles
		to the LLVM code \subref{subfig:llvm_source} where the two
		\texttt{mul} instructions (i.e., \texttt{c0} and \texttt{c1})
		are incompatible for vectorization since \texttt{c0} is used
		before the definition of \texttt{c1}.
		SILVIA rearranges the code \subref{subfig:llvm_rearrange} to
		make \texttt{c0} and \texttt{c1} compatible, by moving the uses
		of \texttt{c0} \acf*{ALAP} while preserving the functionality.
		Finally, SILVIA replaces the two \texttt{mul} instructions with
		a function call to the corresponding \acs*{DSP}-packed
		implementation \subref{subfig:llvm_packed}.}
		\label{fig:inst_rearrange}
	\Description{LLVM IR code snippets at different steps in the SILVIA
	flow.
	}
\end{figure}

For instance, the previously discussed design example defined in
\cref{subfig:c_source}, which computes two parallel multiplications via loop
unrolling, is compiled to the LLVM \ac{IR} in \cref{subfig:llvm_source}.
The first use of the first multiplication (i.e., the \texttt{store} of
\texttt{c0}) comes before the last definition of the operands of the second
multiplication (i.e., the \texttt{load} of \texttt{a1}).
In this scenario, there is no room for placing the \textcolor{revmod}{packed}
multiplication, since the last definition -- first use intervals of the two
multiplications (highlighted by the arrows in \cref{subfig:llvm_source}) do not
intersect.

To maximize the room for \textcolor{revmod}{\ac{DSP}-packed} calls, the
\texttt{moveUsesALAP} function moves the candidate's uses \ac{ALAP} within
\iac{BB}.
SILVIA preserves the data dependencies by analyzing the definition--uses chains
and exploiting the memory aliasing analysis LLVM infrastructure.
Moreover, it conservatively assumes that function calls may alias with other
function calls or memory operations, to guarantee the functionality without
expensive inter-procedural analysis outside the \ac{BB} scope.

When building the tuples of candidates to pack, SILVIA checks if there exists a
\emph{\textcolor{revmod}{packed} insertion point}.
The function tests if the last definition -- first use interval of a candidate
intersects with the interval of a tuple (i.e., they intersect when the last
definition of the candidate comes before the first use of the tuple and vice
versa).
It is notable that the existence of a valid insertion point implies the tuple
is free from interdependencies between its candidates.

\subsubsection{Operation-specific tuple validity}\label{subsubsec:can_pack}
The \texttt{canPack} virtual function exposed by SILVIA is an additional hook
for filtering the candidates to be added to a tuple, based on the requirements
of the specific \textcolor{revmod}{packed} operation.

The \texttt{SILVIAMuladd} class overrides the \texttt{canPack} function to
check whether a candidate shares one of its operand with the other candidates
in the tuple, in compliance with \cref{eqn:packing,eqn:4b_mul}.

The \texttt{SILVIAAdd}'s \texttt{canPack} function does not perform any further
check, since \iac{SIMD} \ac{DSP} can compute any tuple of independent additions.

\subsection{Tuple packing}\label{subsec:packing}
In the \texttt{packTuple} function, SILVIA processes
\textcolor{revmod}{packed} tuple and
creates a call to the function implementing the
\textcolor{revmod}{\ac{DSP}-packed} functionality in a
valid \textcolor{revmod}{packed} insertion point (as explained in
\cref{subsubsec:insert_point}), generating, for instance, the code in
\cref{subfig:llvm_packed}.

The called function can:
\begin{itemize}
\item Directly implement the optimized \textcolor{revmod}{\ac{DSP}-packed}
module in the LLVM \ac{IR}.
For instance, \texttt{SILVIAMuladd} specifies the operations to be computed by
the \ac{SIMD} \ac{DSP} such that the \ac{HLS} tool maps them to a single
\ac{DSP}.
\item Be a functionally-equivalent placeholder, such that the \ac{HLS} tool
generates a dedicated module with the desired functionality and interface which
can be replaced with a custom \ac{RTL} module in the following step, described
in \cref{subsec:replacement}.
For example, the \ac{SIMD} adder requires setting the \texttt{use\_simd} Vivado
synthesis directive that is not controllable from the LLVM \ac{IR}; the
multiplication between four signed 4-bit operands and one shared 4-bit operand
is implemented at \ac{RTL}~\cite{finn}.
\end{itemize}

The \texttt{packTuple} step totally depends on the specific
\textcolor{revmod}{packed} operation.
For instance, the \texttt{SILVIAMuladd} pass ensures that \cref{eqn:mad_chain}
is satisfied. In case the tuple contains more than $N$ pairs of candidates,
SILVIA splits them into multiple balanced \ac{DSP} chains and sums the
remaining partial \acp{MAD} with an external adder tree to avoid overflow.

\subsection{Tuple replacement}\label{subsec:replacement}
Given a packed tuple, SILVIA replaces the uses of the tuple with the values
computed by the \textcolor{revmod}{packed} function, such that the original
tuple becomes dead code (i.e., without any use), and calls the \emph{dead code
elimination} LLVM pass to remove the leftover original tuple.

Finally, SILVIA replaces the \ac{HLS}-generated \ac{RTL} modules corresponding
to the placeholder functions called by \texttt{packTuple} with the related
custom \textcolor{revmod}{\ac{DSP}-packed} \ac{RTL} modules.
Vitis HLS provides the ``black box'' functionality, that is similar but
currently has too many limitations to be used for this purpose.
Therefore, SILVIA re-implemented it.

\subsection{Impact on the HLS backend}\label{subsec:be_impact}
Replacing tuples of instructions with their packed counterpart impacts the
behavior of the \ac{BE} during scheduling (i.e., the assignment of the
execution of each operation to specific clock cycles) and  resource sharing
(i.e., the binding of different operations to the same functional unit in
different clock cycles).

\subsubsection{Impact on scheduling}
The pipeline \ac{II} (i.e., the number of clock cycles between the start of
successive pipeline iterations) is a key parameter of a schedule, since the
throughput is inversely proportional to it.
The \ac{DDG} (i.e., the graph whose nodes are the instructions in a design and
whose edges are the data dependencies between the instructions) enables to
compute the minimum \ac{II} allowed by the data dependencies as
\begin{equation}
II_{\min} \triangleq \max\limits_\theta
	\left\lceil
	\dfrac{\mathrm{latency}_\theta}{\mathrm{distance}_\theta}
	\right\rceil,
	\label{eqn:min_ii}
\end{equation}
where $\theta$ is any cycle in the \ac{DDG}, $\mathrm{latency}_\theta$ is the sum of
the latencies of each node belonging to $\theta$ and $\mathrm{distance}_\theta$
is the total dependence distance along each edge belonging to
$\theta$~\cite{sw_pipeline}.
The cycle maximizing \cref{eqn:min_ii} is called the \emph{critical cycle}.

\begin{figure}
	\subcaptionbox{Source code.\label{subfig:critical_cycle_src}}{
		\includegraphics{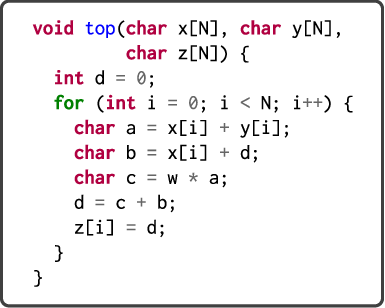}
	}
	\hfil
	\subcaptionbox{Original \acs*{DDG}.\label{subfig:critical_cycle_ddg_orig}}{
		\includegraphics{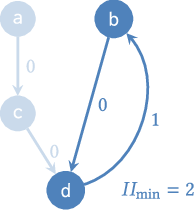}
	}
	\hfil
	\subcaptionbox{Packed \acs*{DDG}.\label{subfig:critical_cycle_ddg_pack}}{
		\includegraphics{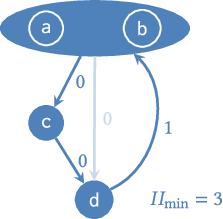}
	}
	\caption{Example of an edge-case design where packing multiple
		operations to the same \acs*{DSP} is detrimental to the
		\acf*{II} of the pipeline.
		The \acf*{DDG}~\subref{subfig:critical_cycle_ddg_orig}
		corresponding to the original source
		code~\subref{subfig:critical_cycle_src}, where the nodes are
		the instructions and the edges are the data dependencies
		between the instructions labeled with their distance, has a
		critical cycle (highlighted in dark blue) which determines a
		minimum \acs*{II} of 2 clock cycles (i.e., the maximum ceiled
		ratio between the total latency and the total distance along
		any cycle in the \acs*{DDG}), assuming a latency of 1 clock
		cycle for each operation.
		Packing \texttt{a} and \texttt{b} to the same \acs*{DSP}
		introduces a new critical cycle in the
		\acs*{DDG}~\subref{subfig:critical_cycle_ddg_pack} that
		increases the minimum \acs*{II} to 3 clock cycles.
		}
	\label{fig:critical_cycle}
	\Description{The effects of SILVIA on a sample data dependence graph.}
\end{figure}

\Cref{fig:critical_cycle}
shows an example where packing increases the minimum \ac{II}.
The packing changes the original \ac{DDG}
(\cref{subfig:critical_cycle_ddg_orig}), corresponding to the source code in
\cref{subfig:critical_cycle_src}, to the \ac{DDG} in
\cref{subfig:critical_cycle_ddg_pack}, where the candidates belonging to the
packed tuple are merged into a single super-node, introducing a new critical
cycle, due to the fictitious data dependencies between the candidates of the
tuple.

There are multiple possible solutions to overcome this issue, such as
discarding the candidates belonging to \ac{DDG} cycles, which is
computationally efficient but might be too conservative.
Another solution is to drop the tuples that introduce new critical cycles,
which is optimal but requires information that might be unavailable before the
\ac{BE} stage (e.g., the latency of the operations).
However, since the depicted scenario empirically proved to be uncommon (e.g.,
none of the several benchmarks considered in \cref{sec:results} suffered from
it), the management of this corner case is left to future work.

\subsubsection{Impact on resource sharing}
Resource sharing applies to operations of the same type only.
For instance, in the \texttt{SILVIAAdd} case, if an addition instruction is
compatible with the additions belonging to a packed tuple, the scheduler would
fail to share the same functional unit between the addition and the tuple,
because the \ac{BE} would recognize them as different operations.
To overcome this issue, SILVIA could map the addition instruction to a packed
\ac{DSP}, even if it is the only instruction in the tuple. With this approach
both the tuple and the unpaired addition would be recognized as operations of
the same type and the resource sharing would successfully apply.
Also this optimization is left to future work.

\section{Results}\label{sec:results}
\begin{figure}
	\includegraphics{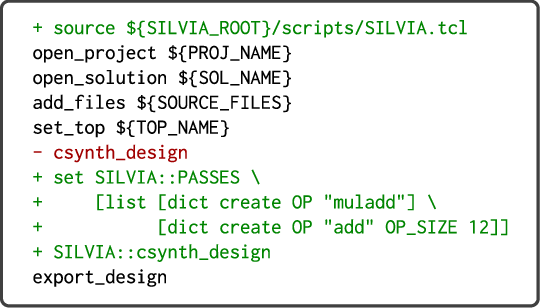}
	\caption{Modifications to the Vitis HLS synthesis script for executing
	the optimized SILVIA flow. Users just need to update the Vitis HLS
	synthesis script by specifying which SILVIA passes to run, via the
	\texttt{SILVIA::PASSES} list, and running the custom
	\texttt{SILVIA::csynth\_design} command.}
    \label{fig:csynth}
    \Description{Line-by-line differences between a standard Vitis HLS
    synthesis script and a SILVIA-optimized one.}
\end{figure}

SILVIA is integrated in \textcolor{revadd}{the flow of} the AMD Vitis HLS
2023.2 tool. To launch the SILVIA optimized flow (\cref{fig:workflow}),
designers just need to update the synthesis script by replacing the standard
synthesis command (i.e., \texttt{csynth\_design}) with the SILVIA's custom one
(i.e., \texttt{SILVIA::csynth\_design}), after selecting the SILVIA passes to
run (i.e., by populating the \texttt{SILVIA::PASSES} variable with the ordered
list of passes), as shown in \cref{fig:csynth}.
{\color{revadd}
The \texttt{OP} option selects the SILVIA pass (e.g., \texttt{"add"}
for the \texttt{SILVIAAdd}) and the \texttt{OP\_SIZE} option specifies
the maximum size of the operations to pack (e.g., \texttt{12} or \texttt{24}
bits for the \texttt{SILVIAAdd}).
Additionally, there are pass-specific options.
\texttt{SILVIAAdd} provides the \texttt{INST} option to pack either addition or
subtraction operations.
\texttt{SILVIAMuladdd} exposes the \texttt{MAX\_CHAIN\_LEN} option to limit the
length of cascaded \ac{DSP} chains.
By default, \texttt{SILVIAMuladdd} chains up to $N$ \acp{DSP}, where $N$ is
defined as \cref{eqn:mad_chain}.
Longer chains save more logic resources, summing the partial products with the
\ac{DSP}'s internal post-adder, but consume more memory resources, due the
deeper pipelines.
Therefore, the \texttt{MAX\_CHAIN\_LEN} option enables trading off between
logic and memory resources.
}

SILVIA automatically identifies and packs the parallel additions,
multiplications, and \acp{MAD} naturally exposed by \ac{HLS} designs on a set
of diverse benchmarks, ranging from simple \ac{BLAS} accelerators to complex
high-performance open-source designs, as discussed in
\cref{subsec:generic_bench}.

\Cref{subsec:cnn_benchs} analyzes the use case of \ac{CNN} accelerators,
specifically the \ac{SOTA} frameworks NN2FPGA~\cite{minnella2023design} and
FINN~\cite{finn}.
These frameworks implement the \ac{MAD} packing supported by
\texttt{SILVIAMuladdd} in a user-directed manner, while SILVIA is fully
automated, allowing to compare the \ac{QoR} of SILVIA with the
manually-optimized designs.

\subsection{General purpose benchmarks}\label{subsec:generic_bench}
The benchmarks are groupable into two main categories,
\emph{addition-intensive benchmarks}, to evaluate the \texttt{SILVIAAdd} pass
\textcolor{revadd}{(configuring the \texttt{SILVIA::PASSES} variable to pack
the 12-bit additions and the 24-bit additions)}, and \emph{multiplication and
\ac{MAD}-intensive benchmarks}, to evaluate the \texttt{SILVIAMuladd} pass
\textcolor{revadd}{(configuring the \texttt{SILVIA::PASSES} variable to pack
the 4-bit multiplications and the 8-bit multiplications/\acp{MAD} with a
maximum \ac{DSP} chain length of \num{3})}.

The addition-intensive benchmarks include:
\begin{itemize}
    \item The vector addition from the Xilinx example
        designs~\cite{xilinx2024hlsintroductoryexamples}, summing two vectors
        of 192 8-bit elements.
    \item \Iac{SNN} convolutional layer~\cite{ottati2024efficient} accelerator,
        with a $24\times24\times64$ input feature map and $3\times3\times64\times128$ filter.
\end{itemize}

The multiplication-intensive benchmarks include:
\begin{itemize}
    \item Four \ac{BLAS} accelerators, namely the \ac{MVM} between a
        $192\times192$ matrix and a $192\times1$ vector, the \ac{MMM} between
        two $192\times192$ matrices, and the \texttt{axpy} and \texttt{scal}
        kernels on 512-element vectors from the Vitis
        Libraries~\cite{xilinx2024}. In every case, the inputs are 
	8-bit integers. The \ac{MMM} also includes a 4-bit unsigned integers
	configuration.
    \item The \ac{GSM} kernel from the CHstone benchmark
        suite~\cite{hara2008chstone}, with 8-bit words.
    \item The forward 3D \acl{HBC} \ac{RTM} accelerator from the Vitis
        Libraries~\cite{xilinx2024}, running on 8-bit fixed-precision data.
    \item The \ac{GAT} \acl*{GNN} accelerator from FlowGNN~\cite{flowgnn}, with
        8-bit fixed-point data.
\end{itemize}

{\color{revmod}
Each benchmark is implemented in different versions, including:
\begin{itemize}
\item the baseline \ac{DSP} (BD) designs, generated with the standard Vitis HLS
flow, configured to bind the operations of interest (i.e., additions in the
addition-intensive benchmarks and multiplications in the
multiplication-intensive benchmarks) to \acp{DSP}, via the \texttt{config\_op}
command, for more direct comparison,
{\color{revadd}\item the baseline unconstrained (BU) version, generated with
the standard Vitis HLS flow without any resource binding constraint, and
}
\item the designs optimized with the SILVIA (S) flow.
\end{itemize}
}

{\color{revmod}
\Cref{tab:benchs_add,tab:benchs_mad}
\begin{table}
{\color{revmod}
	\caption{\color{revmod}\Acl*{PPA} of the benchmarks.
    The baseline \acs*{DSP}~(BD) and baseline unconstrained~(BU) results are
    obtained with the standard Vitis HLS flow.
    The SILVIA~(S) results originate from the optimized SILVIA flow.
    Additionally, the BD and S designs forced Vitis HLS to bind the
    addition or multiplication operations to \acsp*{DSP}.
    The ``\textleftarrow'' signifies that Vitis HLS generated the same design
    for both BU and BD versions (i.e., it automatically bound every
    multiplication to \iac{DSP}).
    \emph{Ops/Unit} is the average number of operations mapped to a single
    arithmetic unit.
    The area and power data derive from the reports of the Vivado
    implementation, with a clock frequency constraint of \SI{250}{\mega\hertz}
    and the default synthesis and implementation settings.
    The maximum clock frequency is the highest at which post-routing timing is
    met, using steps of \SI{25}{\mega\hertz}.
    The on-chip memory utilization is omitted because SILVIA does not impact it.
 }
 \subcaptionbox{Addition-intensive benchmarks.\label{tab:benchs_add}}{
    \footnotesize
    \addtolength{\tabcolsep}{-2.5pt}
    {
	\begin{tabular}{c*{2}{S[table-format=1.2]}*{3}{S[table-format=4.0]}*{3}{*{3}{S[table-format=2.2]}}*{3}{S[table-format=4.0]}*{3}{S[table-format=3.0]}}
		\toprule
		& \multicolumn{2}{c}{\textbf{Ops/Unit} (\si{\one})} & \multicolumn{3}{c}{\textbf{DSP} (\si{\one})} & \multicolumn{3}{c}{\textbf{Logic LUT} (\si{\kilo\relax})} & \multicolumn{3}{c}{\textbf{Mem.\ LUT} (\si{\kilo\relax})} & \multicolumn{3}{c}{\textbf{FF} (\si{\kilo\relax})} & \multicolumn{3}{c}{\textbf{Power} (\si{\milli\watt})} & \multicolumn{3}{c}{$\mathbf{F_{clk}^{max}}$ (\si{\mega\hertz})} \\
		\cmidrule(lr){2-3}
		\cmidrule(lr){4-6}
		\cmidrule(lr){7-9}
		\cmidrule(lr){10-12}
		\cmidrule(lr){13-15}
		\cmidrule(lr){16-18}
		\cmidrule(lr){19-21}
		\multirow{-2}{*}[.3em]{\textbf{Bench.}} & {BD,BU} & {S} & {BD} & {BU} & {S} &
  {BD} & {BU} & {S} & {BD} & {BU} & {S} & {BD} & {BU} & {S} & {BD} & {BU} & {S} & {BD} & {BU} & {S} \\
		\midrule
        vadd~\cite{xilinx2024hlsintroductoryexamples} & 1.00 & 3.40 & 68 & 0 & 20 & 3.49 & 3.53 & 3.30 & 0.59 & 0.59 & 0.59 & 7.80 & 8.18 & 7.29 & 365 & 359 & 345 & 475 & 475 & 425 \\
        SNN~\cite{ottati2024efficient} & 1.00 & 3.19 & 150 & 0 & 47 & 1.47 & 2.05 & 1.19 & 0.00 & 0.00 & 0.05 & 2.72 & 3.85 & 2.95 & 439 & 352 & 373 & 450 & 600 & 450 \\
	\midrule
        N.\ gmean & {1.00} & {\textbf{3.29}} & {1.00} & {0.00} & {\textbf{0.30}} & {1.00} & {1.19} & {0.87} & {1.00} & {1.00} & {7.07} & {1.00} & {1.22} & {1.01} & {1.00} & {0.89} & {0.90} & {1.0} & {1.15} & {0.95} \\
		\bottomrule
	\end{tabular}
	}
	}
	}

	\bigskip
	{\color{revmod}
	\subcaptionbox{Multiplication-intensive benchmarks.\label{tab:benchs_mad}}{
    \footnotesize
    \addtolength{\tabcolsep}{-2.5pt}
    {
	\begin{tabular}{c*{2}{S[table-format=1.2]}*{3}{S[table-format=4.0]}*{3}{*{3}{S[table-format=2.2]}}*{3}{S[table-format=4.0]}*{3}{S[table-format=3.0]}}
		\toprule
		& \multicolumn{2}{c}{\textbf{Ops/Unit} (\si{\one})} & \multicolumn{3}{c}{\textbf{DSP} (\si{\one})} & \multicolumn{3}{c}{\textbf{Logic LUT} (\si{\kilo\relax})} & \multicolumn{3}{c}{\textbf{Mem.\ LUT} (\si{\kilo\relax})} & \multicolumn{3}{c}{\textbf{FF} (\si{\kilo\relax})} & \multicolumn{3}{c}{\textbf{Power} (\si{\milli\watt})} & \multicolumn{3}{c}{$\mathbf{F_{clk}^{max}}$ (\si{\mega\hertz})} \\
		\cmidrule(lr){2-3}
		\cmidrule(lr){4-6}
		\cmidrule(lr){7-9}
		\cmidrule(lr){10-12}
		\cmidrule(lr){13-15}
		\cmidrule(lr){16-18}
		\cmidrule(lr){19-21}
		\multirow{-2}{*}[.3em]{\textbf{Bench.}} & {BD,BU} & {S} & {BD} & {BU} & {S} &
  {BD} & {BU} & {S} & {BD} & {BU} & {S} & {BD} & {BU} & {S} & {BD} & {BU} & {S} & {BD} & {BU} & {S} \\
		\midrule
        MVM & 1.00 & 2.00 & 64 & {\textleftarrow} & 32 & 1.57 & {\textleftarrow} & 1.42 & 0.68 & {\textleftarrow} & 0.68 & 1.65 & {\textleftarrow} & 2.42 & 395 & {\textleftarrow} & 389 & 375 & {\textleftarrow} & 400 \\
        MMM & 1.00 & 2.00 & 64 & {\textleftarrow} & 32 & 1.60 & {\textleftarrow} & 1.62 & 0.46 & {\textleftarrow} & 0.59 & 1.68 & {\textleftarrow} & 2.42 & 441 & {\textleftarrow} & 459 & 350 & {\textleftarrow} & 350 \\
        MMM-4b & 1.00 & 4.00 & 64 & {\textleftarrow} & 16 & 1.50 & {\textleftarrow} & 2.00 & 0.23 & {\textleftarrow} & 0.27 & 1.26 & {\textleftarrow} & 2.11 & 438 & {\textleftarrow} & 440 & 300 & {\textleftarrow} & 350 \\
	\texttt{scal}~\cite{xilinx2024} & 1.00 & 2.00 & 64 & 0 & 32 & 2.46 & 4.54 & 2.46 & 0.01 & 0.01 & 0.01 & 6.85 & 8.44 & 7.36 & 354 & 411 & 351 & 475 & 475 & 475 \\
        \texttt{axpy}~\cite{xilinx2024} & 1.00 & 2.00 & 64 & {\textleftarrow} & 32 & 4.07 & {\textleftarrow} & 4.57 & 0.01 & {\textleftarrow} & 0.27 & 13.58 & {\textleftarrow} & 14.10 & 486 & {\textleftarrow} & 495 & 450 & {\textleftarrow} & 375 \\ 
        \acs*{GSM}~\cite{hara2008chstone} & 1.00 & 1.58 & 41 & {\textleftarrow} & 24 & 0.80 & {\textleftarrow} & 0.98 & 0.03 & {\textleftarrow} & 0.04 & 0.78 & {\textleftarrow} & 1.33 & 333 & {\textleftarrow} & 324 & 350 & {\textleftarrow} & 350 \\
        \acs*{RTM}~\cite{xilinx2024} & 1.00 & 1.14 & 274 & 139 & 232 & 20.38 & 25.23 & 18.45 & 5.40 & 5.33 & 5.57 & 29.11 & 31.14 & 29.13 & 966 & 987 & 586 & 200 & 225 & 275 \\
        \acs*{GAT}~\cite{flowgnn} & 1.00 & 1.97 & 1508 & 540 & 768 & 25.16 & 62.17 & 32.75 & 31.07 & 31.07 & 18.02 & 40.19 & 56.80 & 56.94 & 4578 & 4303 & 3208 & 325 & 325 & 325 \\
        \midrule
        N.\ gmean & {1.00} & {\textbf{1.97}} & {1.00} & {0.00} & {\textbf{0.50}} & {1.00} & {1.24} & {1.09} & {1.00} & {1.00} & {1.54} & {1.00} & {1.08} & {1.33} & {1.00} & {1.01} & {0.92} & {1.00} & {1.01} & {1.05} \\
		\bottomrule
	\end{tabular}
	}
	}
}
\end{table}
report the \acl*{PPA} of the benchmarks, collected from the post-implementation
reports of AMD Vivado 2023.2, targeting UltraScale \ac{FPGA} boards
(specifically, the AMD Kria KV260, except for the \ac{GAT} benchmark, which
targets the AMD ZCU102, due to its higher resource requirements), with a clock
constraint of \SI{250}{\mega\hertz} (except for the \ac{RTM} benchmark,
constrained to \SI{200}{\mega\hertz} due to the timing critical path of its BD
and BU versions).
The maximum clock frequency is the highest at which post-routing timing is met,
at a granularity of \SI{25}{\mega\hertz}.
The operation density (Ops/Unit) is defined as the ratio between the number of
arithmetic operations and the number of functional units computing them, at the
\ac{IR} level.
}

SILVIA increases the mean operation density to \num{3} for the addition
benchmarks and to \num{2} for the multiplication ones, by automatically
identifying and packing the compatible operations available in the benchmarks
to \acp{DSP}, \textcolor{revadd}{without affecting the cycle count  performance}.
{\color{revmod}These results demonstrate that SILVIA successfully
achieves its goal of \emph{automating \ac{DSP} packing} within the
\ac{HLS} flow.
Although the effectiveness of \ac{DSP} packing itself is well-established, as
evidenced by numerous \ac{SOTA}
designs~\cite{minnella2023design,li2023firefly,zhang2022wsq,deepburning}
exploiting it, the rest of this section analyzes its impact on \acl*{PPA}
metrics.
}

\begin{figure}
\captionsetup[subfigure]{justification=centering}
	\subcaptionbox{Addition-intensive benchmarks.\label{subfig:add_saving}}{
		\includegraphics{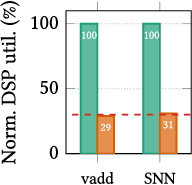}
	}
	\hfil
	\subcaptionbox{Multiplication-intensive benchmarks.\label{subfig:mul_saving}}{
		\includegraphics{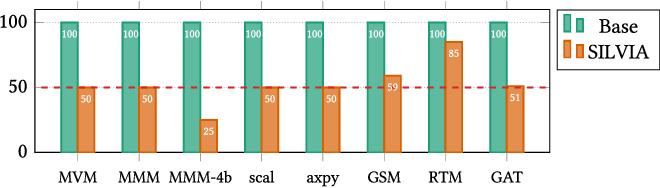}
	}
	\caption{The \ac*{DSP} utilization of the different benchmarks,
	normalized on the baseline \ac{DSP} utilization.
	The dashed horizontal line represents the geometric mean of the
	\ac{DSP} utilization of the SILVIA-optimized designs.}
	\label{fig:savings}
	\Description{Bar plots of DSP utilization of general purpose benchmarks.
	The SILVIA-optimized design on average consume 70\% and 50\% less DSPs
	for the addition-intensive and multiplication-intensive benchmarks,
	respectively.}
\end{figure}

In principle, the \ac{DSP} saving should be directly proportional to the
operation density.
However, the \ac{HLS} and the logic synthesis reduce the \ac{DSP} utilization
of the baseline of some benchmarks too (i.e., the used \acp{DSP} are less than
the total number of operations), thanks to different optimizations such as
resource sharing or three-input additions mapping to a single \ac{DSP},
slightly reducing the advantage of SILVIA over the baseline.
Nevertheless, on average, SILVIA saves \SI{70}{\percent} \acp{DSP} in the
addition benchmarks and \SI{50}{\percent} \acp{DSP} in the multiplication
benchmarks with respect to the baseline \ac{DSP}.
Those results suggest that the wider space available when SILVIA optimizes the
\ac{DSP} utilization at higher abstraction levels (i.e., at the LLVM \ac{IR}
level, rather than at the \ac{RTL}) enables achieving better \ac{QoR}.
\Cref{fig:savings}
summarizes the \ac{DSP} utilization of each benchmark, normalized over the
baseline \ac{DSP} utilization.

The \ac{II} of the pipelines of each benchmark is the same in both the baseline
and the SILVIA-optimized designs (i.e., the scenario depicted in
\cref{subsec:be_impact} is never encountered). Therefore, SILVIA does not affect
the throughput of the designs.

{\color{revmod}
\Ac{DSP} packing introduces data dependencies between originally-independent
instructions, increasing pipelines depth (SILVIA-optimized pipelines are, on
average, \SI{27}{\percent} deeper than baseline).
While deeper pipelining does not impact throughput, which depends on the
\ac{II}, it requires more pipelining registers.
Consequently, SILVIA-optimized designs utilize more memory \acp{LUT} and
\acp{FF}.
}

{
\color{revadd}
\Ac{MAD} packing introduces an average \SI{9}{\percent} overhead in logic
\acp{LUT} due to diverse reasons, including the error-correction logic required
by the \ac{MAD} packing methodologies, and the missed opportunities to compute
\acp{MAD} on a single \ac{DSP} when the additions do not comply with
\cref{eqn:packing}.
For instance, the \texttt{axpy} benchmark adds a third operand $d_i$ to the
products (i.e., $p_i = a_i \cdot c_i + d_i$), whereas the packed \ac{MAD} can
only sum together \ac{DSP}-packed products (i.e., $p = a_i \cdot c_i + a_{i+1}
\cdot c_{i+1}$). Therefore, the SILVIA \texttt{axpy} adds $d_i$ with \ac{LUT}
adders, while the baseline computes one \ac{MAD} per \ac{DSP}.

On average, SILVIA-optimized designs consume \SI{8}{\percent} less power in
multiplication-intensive benchmarks and \SI{10}{\percent} less in
addition-intensive benchmarks compared to the baseline, due to substantial
reductions in \ac{DSP} utilization.

The impact of \ac{DSP} packing on the timing critical path is not uniform.
Some SILVIA-optimized designs achieve higher clock frequencies, likely due to
reduced resource usage easing placement, while others experience lower
frequencies, potentially due to increased routing congestion from higher
computation density in packed \acp{DSP}.
}

SILVIA always executes in less than one second when optimizing each benchmark,
except for the largest designs (i.e., \ac{RTM} and \ac{GAT} require around 6
seconds and 2 minutes, respectively).
In every case, the execution time of SILVIA is negligible with respect to the
whole \ac{HLS} (taking minutes) and implementation (taking from minutes to
hours) flows.

\subsection{CNN acceleration case study}\label{subsec:cnn_benchs}
NN2FPGA~\cite{minnella2023design} and FINN~\cite{finn} are open-source
frameworks that generate the hardware description of \ac{CNN} inference
accelerators.
These frameworks support the \ac{MAD} packing supported by SILVIA (FINN
supports both the factor-2 and the factor-4 packing; NN2FPGA the factor-2
only) in a user-controllable manner. Therefore, they are ideal for comparing
the \ac{QoR} of the manually-optimized designs (i.e., with the packing enabled)
with the automatically SILVIA-optimized ones (i.e., with the packing disabled
and optimized with the SILVIA flow).


\Cref{tab:cnn}
\begin{table}
	\caption{\Acl*{CNN} accelerators results.
    The resource utilization derives from Vivado post-implementation reports.
    The baseline results (B) are obtained synthesizing the designs without
    packing with the standard Vitis HLS flow, the manually-optimized (M) ones
    synthesizing the designs with manual packing with the standard Vitis HLS
    flow, and the SILVIA (S) ones synthesizing the designs without packing
    with the optimized SILVIA flow.
    The clock frequency constraint is \SI{200}{\mega\hertz} for every design
    and timing is always met.
    The on-chip memory utilization is not reported because SILVIA does not
    impact it.
 }
	\label{tab:cnn}
 \centering
 \footnotesize
 {\color{revmod}
	\subcaptionbox{NN2FPGA accelerators.}{
	\begin{tabular}{cc*{3}{S[table-format=3.0]}*{2}{*{3}{S[table-format=2.1]}}*{3}{S[table-format=3.1]}*{3}{S[table-format=2.2]}}
		\toprule
		&& \multicolumn{3}{c}{\textbf{\acs*{DSP}} (\si{\one})} & \multicolumn{3}{c}{\textbf{Logic \acs*{LUT}} (\si{\kilo\relax})} & \multicolumn{3}{c}{\textbf{Mem.\ \acs*{LUT}} (\si{\kilo\relax})} & \multicolumn{3}{c}{\textbf{\acs*{FF}} (\si{\kilo\relax})} & \multicolumn{3}{c}{\textbf{Throughput} (kFPS)}\\
		\cmidrule(lr){3-5}
		\cmidrule(lr){6-8}
		\cmidrule(lr){9-11}
		\cmidrule(lr){12-14}
		\cmidrule(lr){15-17}
		\multirow{-2}{*}[.3em]{\textbf{Goal}} & \multirow{-2}{*}[.3em]{\textbf{Model}} & {B} & {M} & {S} & {B} & {M} & {S} & {B} & {M} & {S} & {B} & {M} & {S} & {B} & {M} & {S} \\
		\midrule
        & ResNet20 & 635 & 318 & 318 & 43.7 & 47.3 & 43.6 & 10.2 & 9.9 & 9.9 & 53.5 & 54.2 & 54.9 & 3.05 & 3.05 & 3.05 \\
        & ResNet8 & 773 & 387 & 387 & 31.9 & 35.4 & 31.7 & 6.0 & 6.0 & 6.3 & 36.0 & 36.2 & 38.8 & 12.20 & 12.20 & 12.20 \\
        \cmidrule(lr){2-17}
        \multirow{-3}{*}[.3em]{\shortstack{Min.\\\acs*{DSP}}} & N.\ gmean & {1.00} & {\textbf{0.50}} & {\textbf{0.50}} & {1.00} & {1.10} & {1.00} & {1.00} & {0.99} & {1.01} & {1.00} & {1.01} & {1.05} & {1.00} & {1.00} & {1.00} \\
        \midrule
        & ResNet20 & 635 & 626 & 626 & 43.7 & 66.7 & 60.3 & 10.2 & 11.4 & 11.5 & 53.5 & 68.7 & 68.1 & 3.05 & 6.10 & 6.10 \\
        & ResNet8 & 773 & 773 & 773 & 31.9 & 61.8 & 53.5 & 6.0 & 8.3 & 8.6 & 36.0 & 63.8 & 64.0 & 12.20 & 24.38 & 24.38 \\
        \cmidrule(lr){2-17}
        \multirow{-3}*[.3em]{\shortstack{Max.\\perf.}} & N.\ gmean & {1.00} & {0.99} & {0.99} & {1.00} & {1.72} & {1.52} & {1.00} & {1.24} & {1.27} & {1.00} & {1.51} & {1.50} & {1.00} & {\textbf{2.00}} & {\textbf{2.00}} \\
		\bottomrule
	\end{tabular}
	}

	\bigskip
	\subcaptionbox{FINN accelerators.}{
	\begin{tabular}{cc*{3}{S[table-format=3.0]}*{2}{*{3}{S[table-format=2.1]}}*{3}{S[table-format=3.1]}*{3}{S[table-format=2.2]}}
		\toprule
		&& \multicolumn{3}{c}{\textbf{\acs*{DSP}} (\si{\one})} & \multicolumn{3}{c}{\textbf{Logic \acs*{LUT}} (\si{\kilo\relax})} & \multicolumn{3}{c}{\textbf{Mem.\ \acs*{LUT}} (\si{\kilo\relax})} & \multicolumn{3}{c}{\textbf{\acs*{FF}} (\si{\kilo\relax})} & \multicolumn{3}{c}{\textbf{Throughput} (kFPS)}\\
		\cmidrule(lr){3-5}
		\cmidrule(lr){6-8}
		\cmidrule(lr){9-11}
		\cmidrule(lr){12-14}
		\cmidrule(lr){15-17}
		\multirow{-2}{*}[.3em]{\textbf{Goal}} & \multirow{-2}{*}[.3em]{\textbf{Model}} & {B} & {M} & {S} & {B} & {M} & {S} & {B} & {M} & {S} & {B} & {M} & {S} & {B} & {M} & {S} \\
		\midrule
        & CNV-8b & 90 & 43 & 43 & 39.4 & 21.1 & 39.4 & 6.6 & 6.8 & 6.6 & 101.7 & 42.4 & 103.5 & 0.05 & 0.05 & 0.05 \\
	& MobileNet-4b & 419 & 163 & 163 & 63.0 & 52.9 & 63.3 & 27.7 & 28.1 & 27.8 & 144.0 & 117.1 & 150.8 & 0.10 & 0.10 & 0.10 \\
        \cmidrule(lr){2-17}
        \multirow{-3}*[.3em]{\shortstack{Min.\\\acs*{DSP}}} & N.\ gmean & {1.00} & {\textbf{0.43}} & {\textbf{0.43}} & {1.00} & {0.67} & {1.00} & {1.00} & {1.02} & {1.00} & {1.00} & {0.58} & {1.03} & {1.00} & {1.00} & {1.00} \\
        \midrule
        & CNV-8b & 90 & 86 & 86 & 39.4 & 22.9 & 39.8 & 6.6 & 8.1 & 7.9 & 101.7 & 44.0 & 106.6 & 0.05 & 0.10 & 0.10 \\
        & MobileNet-4b & 419 & 427 & 427 & 63.0 & 66.7 & 79.7 & 27.7 & 28.3 & 27.8 & 144.0 & 123.4 & 185.0 & 0.10 & 0.26 & 0.27 \\
        \cmidrule(lr){2-17}
        \multirow{-3}*[.3em]{\shortstack{Max.\\perf.}} & N.\ gmean & {1.00} & {0.99} & {0.99} & {1.00} & {0.78} & {1.13} & {1.00} & {1.12} & {1.10} & {1.00} & {0.61} & {1.16} & {1.00} & {\textbf{2.28}} & {\textbf{2.32}} \\
		\bottomrule
	\end{tabular}
	}
	}
\end{table}
shows the post-implementation results at a clock frequency of
\SI{200}{\mega\hertz} generated by Vivado 2022.2 targeting the AMD Kria KV260
for the NN2FPGA designs and Vivado 2023.2 targeting the AMD ZCU102 for the FINN
designs.
The throughput is measured from hardware execution.

{
\color{revmod}The \emph{Minimum \ac{DSP}} experiments are set up as a
throughput-constrained \ac{DSP} minimization problem.
Therefore, both baseline and optimized designs have the same parallelism and
the \ac{DSP} packing minimizes the \ac{DSP} resources without affecting the
throughput.
}
SILVIA successfully matches the \ac{DSP} utilization of the manually optimized
designs, both from NN2FPGA and FINN.

{
\color{revmod}The \emph{Maximum performance} experiments are set up as a
\ac{DSP}-constrained throughput maximization problem.
The higher \ac{DSP} operation density of the optimized designs enables larger
parallelism at the same number of \acp{DSP}, maximizing the throughput.
}
Also in this case, the SILVIA's \ac{QoR} is on par with the manually-optimized
designs, matching their performance per \ac{DSP}.
{
\color{revadd}
As expected, the optimized designs consume more \acp{LUT} and \acp{FF} than the
baseline due to the increased data demands from higher parallelism.
Notably, SILVIA-optimized designs consistently use fewer logic \acp{LUT} than
manually optimized ones, as the \ac{HLS} tool fails to efficiently bind operand
concatenation from NN2FPGA’s source-level packing to the \ac{DSP} pre-adder,
instead mapping it to \iac{LUT}-based adder.
}

In both experiments, the logic \ac{LUT} and \ac{FF} utilizations of the
manually-optimized FINN designs is significantly lower than the SILVIA's ones
because the whole convolutional layer with packing is fine-tuned at \ac{RTL},
rather than using the \ac{HLS} model instantiated when the packing is disabled.



\begin{figure}
	\includegraphics{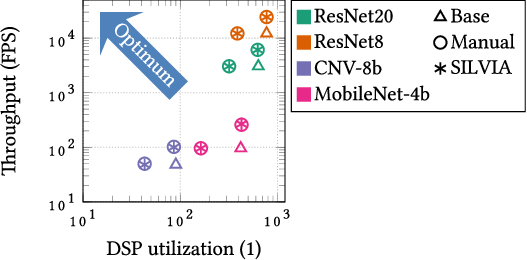}
	\caption{The \acl*{CNN} inference accelerators case study in the
	post-implementation \ac{DSP} utilization versus hardware-measured throughput
	space. SILVIA successfully matches the \acl*{QoR} of the manually-optimized
	designs, both when optimizing for area and for resources.}
	\label{fig:cnn_dsp_perf}
	\Description{CNN inference accelerators in the throughput versus DSP
	utilization space. The SILVIA-optimized designs overlap the
	manually-optimized ones and Pareto-dominate the baseline designs.}
\end{figure}

\Cref{fig:cnn_dsp_perf}
shows that SILVIA effectively matches the manually optimized designs in the
\ac{DSP} versus throughput space, as their design points consistently overlap.
This demonstrates SILVIA's ability to automatically Pareto-dominate the
baseline designs.

\section{Conclusion}
This work proposes SILVIA, the first open-source LLVM infrastructure to automatically
identify and \textcolor{revmod}{optimize \ac{DSP}-packable} operations in
\ac{HLS} \ac{FPGA} designs.
SILVIA can pack four up-to-12-bit or two up-to-24-bit additions or
subtractions, two up-to-8-bit or four up-to-4-bit multiplications, or two
up-to-8-bit \acp{MAD} on a single \ac{DSP}.
Moreover, it is designed to readily support other operations, reusing most of
the existing infrastructure.

SILVIA automatically identifies and optimizes for the
\textcolor{revmod}{superword-level} parallelism a diverse set of benchmarks,
reducing the \ac{DSP} utilization for addition instructions by
\SI{70}{\percent} and for multiplications and \acp{MAD} by \SI{50}{\percent},
on average.
Moreover, SILVIA achieves results comparable with manually optimized \ac{SOTA}
designs.

The experimental results prove that SILVIA is a further step towards higher
abstraction levels for custom hardware acceleration, moving the hardware
knowledge and code analysis from the designer's to the compiler's
responsibility.
This enables \ac{HLS} source code more abstracted from hardware, resulting in
cleaner and more maintainable code bases and making hardware acceleration more
accessible to non-hardware-experts.

\begin{acks}
	This work was partially supported by the Key Digital Technologies
	Joint Undertaking under the REBECCA Project with grant agreement
	number 101097224, receiving support from the European Union,
	Greece, Germany, Netherlands, Spain, Italy, Sweden, Turkey,
	Lithuania, and Switzerland. It was also partially supported by the Spoke 1 on
	Future HPC of the Italian Research Center on High-Performance Computing, Big
	Data and Quantum Computing (ICSC) funded by MUR Mission 4 -- Next Generation
	EU.
\end{acks}

\bibliographystyle{ACM-Reference-Format}
\bibliography{main}


\end{document}